\documentclass[preprint,showpacs,amsmath,amssymb,floatfix,pre]{revtex4}
\usepackage{graphicx}
\begin{document}
\title{Phase diagrams of the transverse Ising antiferromagnet in the presence of the longitudinal magnetic field} 
\author{Minos A. Neto}
\email{minos@pq.cnpq.br}
\affiliation{Departamento de F\'{\i}sica, Universidade Federal do Amazonas, 3000, Japiim,
69077-000, Manaus-AM, Brazil}

\author{J. Ricardo de Sousa}
\email{jsousa@pq.cnpq.br}
\affiliation{Departamento de F\'{\i}sica, Universidade Federal do Amazonas, 3000, Japiim,
69077-000, Manaus-AM, Brazil}
\affiliation{National Institute of Science and Technology for Complex Systems, 3000, Japiim,
69077-000, Manaus-AM, Brazil}

\date{\today}

\begin{abstract}
\textbf{ABSTRACT}

In this paper we study the critical behavior of the two-dimensional antiferromagnetic Ising model in both uniform longitudinal ($H$) and transverse 
($\Omega $) magnetic fields. Using the effective-field theory (EFT) with correlation in single site cluster we calculate the phase diagrams in the 
$H-T$ and $\Omega -T$ planes for the square ($z=4$) lattices. We have only found second order phase transitions for all values of fields and reentrant 
behavior was not observed.

\end{abstract}

\maketitle

\section{Introduction\protect\nolinebreak}
In the last decade there has been an increasing number of works dealing with magnetic models to study quantum phase transitions. In particular, 
considerable interest has been directed to the transverse Ising model (TIM) used to describe a variety of physical systems \cite{deGennes,
stinchcombe,samara}. It was originally introduced by de Gennes \cite{deGennes} as a pseudospin model for hydrogen-bonded ferroelectrics such 
as $KH_{2}PO_{4}$ in the order-disorder phenomenon with tunnelling effects so, it has been successfully used to study a number of problems of 
phase transitions associated with order-disorder phenomena in other systems \cite{blinc,1elliott}. It provides a good description to analyze some 
real anisotropic magnetic materials in a transverse field. 

Theoretically, various methods has been employed to study the criticality of the TIM such as renormalizaton group (RG) method \cite{fisher}, effective 
field theory (EFT) \cite{saber,sarmento1,bouziane}, mean field theory (MFA) \cite{saxena1}, cluster variation method (CVM) \cite{saxena2}, pair 
approximation (PA) \cite{canko}, Monte Carlo (MC) simulations \cite{creswick}, and so on. The critical behavior of the one dimensional TIM has 
already been established through exact results, where the ground-state energy, the elementary excitations and the correlation functions were obtained 
\cite{pfeuty}. The TIM is among the simplest conceivable classes of quantum models in statistical mechanics to study quantum phase transition 
\cite{chakrabarti,sachdev}. The ferromagnetic TIM has been studied intensively. The critical and thermal properties of the ferromagnetic and 
antiferromagnetic TIM are equivalent, but the properties of these two model are very different at not null longitudinal field $(H\neq0)$. For example, 
the ground-state phase transition in the ferromagnetic TIM is smeared out by the longitudinal field in contrast to the antiferromagnetic TIM for which 
the phase diagram remains qualitatively the some at $(H\neq0)$.   

Experimental investigations on the metamagnetic compounds such as $FeBr_{2}$ and $FeCl_{2}$ \cite{vettier} and $Ni(NO_{3})_{2}2H_{2}O$ \cite{sughi1,
sughi2}, under hydrostatic pressure, have been performed. For example, in the $FeBr_{2}$ compound was observed a sharp peak in magnetization measurements under 
a field inclined by $33^{\circ}$ with the $c$ axis (perpendicular to the plane) of the crystal. They concluded that the peak was affected by the ordering 
of the planar spin components. It is obvious that the field can be decomposed into the longitudinal and transverse fields. The model is described by the Ising 
Hamiltonian to which is added a term which represents the effects of the transverse field part. Due to the requisite of non-commutativity of the operators in 
the Hamiltonian, deriving the eigenvalues of the Hamiltonian is a very difficult problem. Therefore, many theoretical methods have been used to investigate 
this system \cite{suzuki,sarmento2,aharony,1ma,2ma}. 

From a theoretical point of view it is known that the effect of the transverse field in the Ising model TIM is to destroy the long-range order of the system. 
Many approximate methods \cite{fisher,saber,sarmento1,bouziane,saxena1,saxena2,canko,creswick,suzuki,sarmento2,aharony,1ma,2ma,kaneyoshi,weng,cassol,dosSantos,
2elliott} have been used to study the critical properties of this quantum model. Some years ago, a simple and versatile scheme, denoted by differential 
operator technique \cite{hk}, was proposed and has been applied exhaustively to study a large variety of problems. In particular, this technique was used to 
treat the criticality of the TIM \cite{barreto} obtaining satisfactory  quantitative results in comparison with more sophisticated methods (for example, MC). 
This method is used in conjugation with a decoupling procedure which ignores all high-order spin correlations (EFT). The EFT included correlations through 
the use of the van der Waerden identity and provided results which are much superior that the MFA. The TIM was first studied by using EFT \cite{barreto} for 
the case of spin $S=1/2$, and generalized \cite{kaneyoshi} for arbitrary spin-$S\geqslant 1/2$. 

On the other hand, the critical and thermal properties of the transverse Ising antiferromagnet in the presence of a longitudinal magnetic field have been few 
studied in the literature \cite{sen,ovchinnikov}. Using the classical approach MFA and the density-matrix renormalization-group method (DMRG) \cite{ovchinnikov}, 
the ground state phase diagram in the $(H-\Omega)$ plane was studied in one-dimensional lattice. Critical line separates the antiferromagnetic (AF) phase 
with long-range order (LRO) from the paramagnetic (P) phase with uniform magnetization. The quantum critical point $\delta_{c}\equiv\left(\frac{\Omega}{J}\right)
_{c}$ decreases as $h_{c}\equiv\left(\frac{H}{J}\right)_{c}$ increase, and is null at $h_{c}=2.0$. The MFA approach does not give the correct qualitative 
description of the critical line \cite{ovchinnikov}. Firstly, the quantum fluctuations shift the ground critical point $\delta_{c}=1.0$ to 
$\delta_{c}=2.0$ at $h=0$ underestimates critical value. Second, the form of the critical line shows an incorrect behavior around of the critical point 
$h_{c}=2.0$. 

By using the EFT approach, Neto and de Sousa \cite{neto04} have studied the ground-state phase diagram of this quantum model on two-dimensional 
(honeycomb $(z=3)$ and square $(z=4)$) lattices, where was discussed the possibility of existence of a reentrant behavior around $h_{c}=z$ critical value. 
The spin correlation effects are partially taken into account in EFT, while it is entirely neglected in MFA. The differences of results given by using EFT 
and MFA show that the spin correlation has important on the phase diagram. The ground-state phase diagram in the $(H-\Omega)$ plane is qualitatively similar to the 
results of Fig. (\ref{estado_fundamental}) in Ref. \cite{ovchinnikov}, but the reentrant behavior found by MFA occurs near $h_{c}=z$.

In the present paper, using EFT we investigate the quantum phase transitions of the Ising antiferromagnetic in both external longitudinal and transverse fields. 
This work is organized as follows: In Sec. II we outline the formalism and its application to the transverse Ising antiferromagnetic in the presence of a 
longitudinal magnetic field; in Sec. III we discuss the results; and finally, in Sec. IV we present our conclusions.

\section{Model and Formalism}

The model studied in this work is the nearest-neighbor ($nn$) Ising antiferromagnet in a mixed transverse and longitudinal magnetic fields divided into two equivalent 
interpenetrating sublattices $A$ and $B$, that is described by the following Hamiltonian

\begin{equation}
H=J\sum\limits_{<i,j>}\sigma _{i}^{z}\sigma_{j}^{z}-H\sum\limits_{i}\sigma _{i}^{z}-\Omega \sum\limits_{i}\sigma_{i}^{x},
\label{1}
\end{equation}%
where $J$ is the AF exchange coupling, $\langle i,j\rangle$ denoted the sum over all pairs of nearest-neighbor spins ($z$) on a $d$-dimensional lattice (here we 
treat the two-dimensional lattices with $z=3$ and $4$), $\sigma _{i}^{\nu }$ is the $\nu (=x,z)$ component of the spin-$1/2$ Pauli operator at site $i$, and $H(\Omega)$ is 
the longitudinal(transverse) magnetic field.

The $\sigma_{i}^{x}$ and $\sigma_{i}^{z}$ spin-$1/2$ Pauli operators do not commute, then a nonzero field $(\Omega)$ transverse to the spin direction causing 
quantum tunneling between the spin-up and spin-down eigenstates of $\sigma_{i}^{z}$ and quantum spin fluctuations. These fluctuations decrease the critical temperature 
$T_{c}$ at which the spins develop long-range order. At a critical field $\Omega_{c}$, $T_{c}$ vanishes, and a quantum phase transition between the AF ordered state and 
a quantum paramagnetic state occur. To the best of our knowledge, the model (\ref{1}) at finite temperature $(T\neq0)$ has not yet been examined in the literature. 
In particular, the ground-state phase diagram was studied by Neto and de Sousa \cite{neto04}, then, in this work we generalize it to analyze the field effects at finite 
temperature by using EFT.  

\begin{figure}[htbp]
\centering
\includegraphics[width=9.6cm,height=5.9cm]{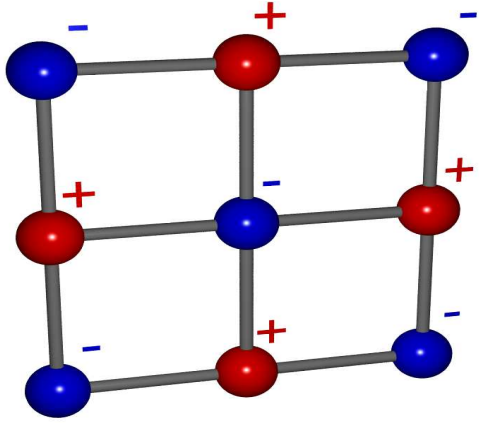}
\caption{Ground state of the quantum Ising antiferromagnet on an square lattice described by Hamiltonian given in Eq. (\ref{1}).} 
\label{estado_fundamental}
\end{figure}

The ground-state of the model (\ref{1}) is characterized by an antiparallel spin orientation in the horizontal and vertical directions and so it exhibits N\'{e}el order 
within the initial sublattices $A$ and $B$ (see Fig. (\ref{estado_fundamental})), that is denoted by the AF state. 

\vspace{0.1cm}

The competition between the antiferromagnetic exchange field presents interesting properties in the phase diagram. In particular, the model (\ref{1}) has an 
AF (ordered) phase in the presence of a field, with the decreasing transition temperature as the fields intensity increases, where at $T=0$ (ground-
state) a second-order transition occurs at both critical field with $H_{c}(=zJ)$ and $\Omega _{c}$.

To treat the model (\ref{1}) on two-dimensional lattices by the EFT approach, we consider a simple example of cluster on a lattice consisting of a central spin and 
$z$ perimeter spins being the nearest-neighbors of the central one. The nearest-neighbor spins are substituted by an effective field produced by the other spins, which 
can be determined by the condition that the thermal average of the central spin is equal to that of its nearest-neighbor ones. The Hamiltonian for this cluster is given 
by

\begin{equation}
\mathcal{H}_{1A}=\left( J\overset{z}{\sum\limits_{\delta }}\sigma
_{(1+\delta)B}^{z}-H\right) \sigma _{1A}^{z}-\Omega \sigma _{1A}^{x},
\label{2}
\end{equation}%
and 
\begin{equation}
\mathcal{H}_{1B}=\left( J\overset{z}{\sum\limits_{\delta }}\sigma
_{(1+\delta )A}^{z}-H\right) \sigma _{1B}^{z}-\Omega \sigma _{1B}^{x},
\label{3}
\end{equation}%
where $A$ and $B$ denote the sublattice.

From the Hamiltonians (\ref{2}) and (\ref{3}), we obtain the average magnetizations in sublattice $A$, $m_{A}=\langle \sigma _{1A}^{z}\rangle$, and $B$, $m_{B}=\langle 
\sigma _{1B}^{z}\rangle$, using the approximate Callen-Suzuki relation \cite{barreto}, that are given by%

\begin{equation}
m_{A}=\left\langle \frac{H-a_{1A}}{\sqrt{(H-a_{1A})^{2}+\Omega ^{2}}}\tanh
\beta \sqrt{(H-a_{1A})^{2}+\Omega ^{2}}\right\rangle,
\label{4}
\end{equation}%
and 
\begin{equation}
m_{B}=\left\langle \frac{H-a_{1B}}{\sqrt{(H-a_{1B})^{2}+\Omega ^{2}}}\tanh
\beta \sqrt{(H-a_{1B})^{2}+\Omega ^{2}}\right\rangle,
\label{5}
\end{equation}%
where $a_{1A}=J\overset{z}{\sum\limits_{\delta }}\sigma_{(1+\delta )B}^{z}$ and $a_{1B}=J\overset{z}{\sum\limits_{\delta }}\sigma _{(1+\delta )A}^{z}$.

Now, using the identity $\exp (\alpha D_{x})F(x)=F(x+a)$ (where $D_{x}=\frac{\partial }{\partial x}$ is the differential operator) and the van der Waerden 
identity for the two-state spin system (i.e., $\exp (a\sigma_{i}^{z})=\cosh (a)+\sigma _{i}^{z}\sinh (a)$) the Eqs. (\ref{4}) and (\ref{5}) are rewritten as 

\begin{equation}
m_{A}=\left\langle \prod_{\delta\neq 0}^{z}(\alpha
_{x}+\sigma _{(1+\delta )B}^{z}\beta _{x})\right\rangle \left.
F(x)\right\vert _{x=0},
\label{6}
\end{equation}%
and
\begin{equation}
m_{B}=\left\langle \prod_{\delta\neq 0}^{z}(\alpha
_{x}+\sigma _{(1+\delta )A}^{z}\beta _{x})\right\rangle \left.
F(x)\right\vert _{x=0},
\label{7}
\end{equation}%
with 
\begin{equation}
F(x)=\frac{H-x}{\sqrt{(H-x)^{2}+\Omega ^{2}}}\tanh \beta \sqrt{%
(H-x)^{2}+\Omega ^{2}},
\label{8}
\end{equation}%
where $\alpha _{x}=\cosh (JD_{x})$ and $\beta _{x}=\sinh (JD_{x})$. The Eqs. (\ref{6}) and (\ref{7}) are expressed in terms of multiple spin correlation functions. 
The problem becomes unmanageable when we try to treat exactly all boundary spin-spin correlation function present in Eqs. (\ref{6}) and (\ref{7}). In this work we 
use a decoupling right-hand sides in Eqs. (\ref{6}) and (\ref{7}), namely

\begin{equation}
\left\langle \sigma_{iA}^{z} \sigma_{jB}^{z} \ldots \sigma_{lA}^{z}\right\rangle \backsimeq m_{A} m_{B}\ldots m_{A}, 
\label{8a}
\end{equation}%
where $i\neq j \neq \ldots \neq l$ and $m_{\mu}=\left\langle \sigma_{i\mu}^{z} \right\rangle $ $(\mu=A,B)$. The approximation (\ref{8a}) neglects correlation 
between different spins but takes relations such as $\left\langle \left( \sigma_{i\mu}^{z} \right)^{2} \right\rangle =1$ exactly into account, while in the usual 
MFA all the self- and multispin correlations are neglected. Using the approximation (\ref{8a}), the Eqs. (\ref{6}) and (\ref{7}) are rewritten by

\begin{equation}
m_{A}=\sum_{p=0}^{z} A_{p}(T_{N},H,\Omega )m_{B}^{p},
\label{9}
\end{equation}%
and 
\begin{equation}
m_{B}=\sum_{p=0}^{z} A_{p}(T_{N},H,\Omega )m_{A}^{p},
\label{10}
\end{equation}%
with

\begin{equation}
A_{p}(T_{N},H,\Omega )=\frac{z!}{p!(z-p)!}\alpha _{x}^{z-p}\beta _{x}^{p}\left.
F(x)\right\vert _{x=0},
\label{11}
\end{equation}%
where the coefficients $A_{p}(T_{N},H,\Omega )$ are obtained by using the relation $\exp (\alpha D_{x})F(x)=F(x+a)$. 

In terms of the uniform $m=\frac{1}{2}(m_{A}+m_{B})$ and staggered $m_{s}=\frac{1}{2}(m_{A}-m_{B})$ magnetizations, and near the critical point we have $m_{s}\rightarrow 0$ 
and $m\rightarrow m_{0}$, the sublattice magnetizaton $m_{A}$ expanded up to linear order in $m_{s}$ (order parameter) is given by 

\begin{equation}
m_{A}=X_{0}(T_{N},H,\Omega ,m_{0})+X_{1}(T_{N},H,\Omega ,m_{0})m_{s},
\label{12}
\end{equation}%
with 
\begin{equation}
X_{0}(T_{N},H,\Omega ,m_{0})=\sum_{p=0}^{z} A_{p}(T_{N},H,\Omega
)m_{0}^{p},
\label{13}
\end{equation}%
and 
\begin{equation}
X_{1}(T_{N},H,\Omega ,m_{0})=-\sum_{p=0}^{z} pA_{p}(T_{N},H,\Omega
)m_{0}^{p-1}. 
\label{14}
\end{equation}

On the other hand, in this work only second-order transitions are observed, therefore, to study the phase diagram only we analyze the Eqs. (\ref{13}) and (\ref{14}) in the limit 
of $m_{\mathbf{s}}\rightarrow 0$\ on can locate the second-order line and using the fact that $m_{A}=m_{0}+m_{s}$ in Eq. (\ref{12}) we obtain 

\begin{equation}
X_{0}(T_{N},h,\delta ,m_{0})=m_{0},
\label{15}
\end{equation}
\begin{equation}
X_{1}(T_{N},h,\delta ,m_{0})=1,
\label{16}
\end{equation}%
at the critical point in which $m_{s}=0$, $\delta \equiv \Omega /J$ and $h\equiv H/J$.

Thus, we can get an analytic solution for second-order transition where $m_{s}$ is the order parameter which is used to describe the phase transition of the model (\ref{1}). 
The magnetizations of two sublattices are not equal for $m_{s}\neq 0$, and the system is in the antiferromagnetic phase. The magnetizations of two sublattices are equal for 
$m_{s}=0$, and the system is in the saturated paramagnetic phase.
   
\section{Results and Discussion}

The numerical determination of the phase boundary (second-order phase transition) is obtained by solving simultaneously Eqs. (\ref{15}) and (\ref{16}). We determined the phase 
diagrams of this quantum model in the $h-T$ and $\delta -T$ planes for the square ($z=4$) lattice that comprises a field-induced AF phase ($m_{s}\neq 0$) at low fields and a 
P phase ($m_{s}=0$) at high fields. In the case $h=0$, we have $m_{A}=-m_{B}$ ($m_{0}=0$) and the critical behavior reduces to the transverse Ising ferromagnetic analyzed in 
Ref. \cite{barreto}. For $z=3$ and $z=4$, the values of the quantum critical point for $h=0$ obtained are $\delta_{c}=1.83$ ($\delta_{c}^{\text{MFA}}=3.00$) and $\delta_{c}=2.75$ 
($\delta_{c}^{\text{MFA}}=4.00$), respectively \cite{neto04}.  In the limit of null fields $h=\delta =0$, we have obtained the value $m_{0}=0$ and $k_{B}T_{N}/J=3.085$ 
(square lattice) and $k_{B}T_{N}/J=2.104$ (honeycomb lattice). The results for the critical temperature obtained by simple EFT with cluster of $N=1$ spin (EFT-1) $k_{B}T_
{N}/J=3.085$, when compared with the exact solution $k_{B}T_{N}/J=2.269\ldots $, for the Ising model on a square lattice, is not quantitatively satisfactory. On the other hand, 
with the increases of the size cluster ($N=2,4,9,\ldots$) we have a possible convergence of the results for the critical temperature. In the present paper we have only 
interested in obtaining qualitative results for the phase diagram. 

The study of quantum phase transition is nowadays one of the main areas of researching in condensed matter physics, so most the experimental and theoretical 
studies are developed to magnetic quantum critical point (QCP). For the square lattice $(z=4)$, the QCP for $h=0$ obtained by using EFT \cite{barreto}, 
$\delta_{c}=2.75$, can be compared with the results found by using other methods, for example, $\delta_{c}=4.00$ of MFA, $\delta_{c}=3.22$ of path-integral Monte 
Carlo simulation \cite{mc}, $\delta_{c}=3.05$ of the DMRG \cite{dm}, $\delta_{c}=3.08$ of the high-temperature series expansion \cite{1elliott}, $\delta_{c}=3.02$ of 
the effective-field renormalization group (EFRG) \cite{efrg}, $\delta_{c}=3.00$ of the PA \cite{canko}, and $\delta_{c}=3.04$ of the cluster MC \cite{mc2}. With the 
increases of the size cluster we will have a possible convergence for the results rigorous of $\delta_{c}$ \cite{dm,1elliott,efrg,canko,mc2}.    

\begin{figure}[htbp]
\centering
\includegraphics[width=8.6cm,height=7.5cm]{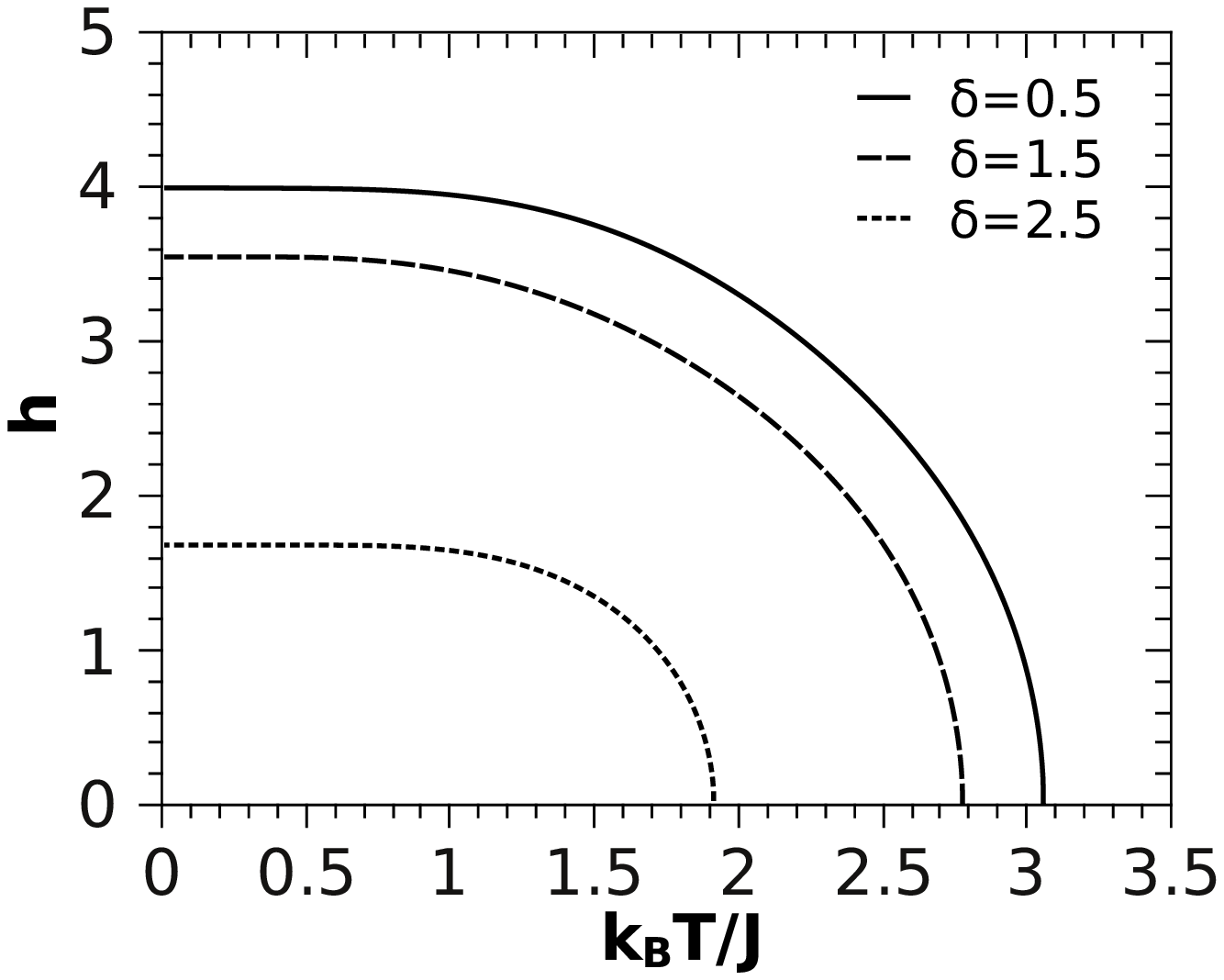}
\caption{Dependence of the reduced critical temperature $k_{B}T/J$ as a function of the reduced magnetic field $h=H/J$ for a square lattice and several 
values of $\delta$. The results are qualitatively identical to the honeycomb lattice.} 
\label{htz4}
\end{figure}

In Fig. (\ref{htz4}), the phase diagram in the $h-T$ plane is presented for the square lattice, with selected values of $\delta$. In this figure, the order of the 
phase transition between the AF and P phases is invariably of second-order for all values of the transverse field. The critical temperature decreases monotonically as 
the longitudinal field $(H)$ increases, and so is null at $h=h_{c}(\delta)$. the critical behavior of $h_{c}$ versus transverse field $(\delta)$ (ground-state phase 
diagram) has been analyzed by Neto and de Sousa \cite{neto04}.

\begin{figure}[htbp]
\centering
\includegraphics[width=8.6cm,height=7.5cm]{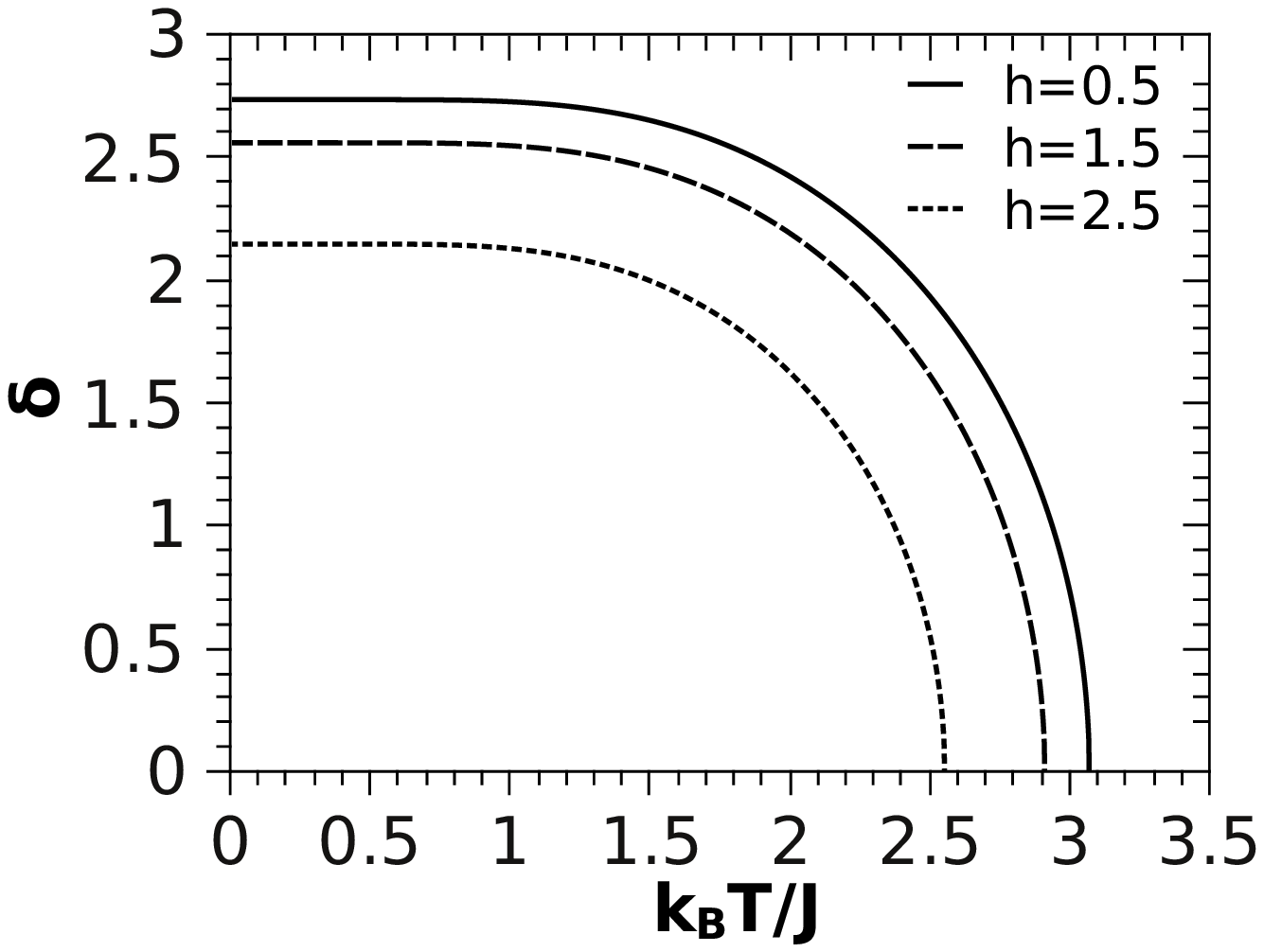}
\caption{Dependence of the reduced critical temperature $k_{B}T/J$ as a function of the reduced magnetic field $\delta =\Omega /J$ for a square lattice and several values 
of $h$. The results are qualitatively identical to the honeycomb lattice.} 
\label{wtz4}
\end{figure}

The same qualitative phase diagram is also observed in the $\delta-T$ plane in Fig. \ref{wtz4} for the square lattice, with selected values of $h$. When the $h$ 
longitudinal field increases the critical curve $T_{c}(\delta)$ decrease. We show that there is no reentrant behavior in the region of low temperature, and also the 
phase transition is of second order for all values of the fields $h$ and $\delta$.

We hope that the present method can be employed for more complex models, for instance, the generalization of the model to treat three-dimensional lattice, where multicritical 
points in the phase diagrams are observed in MFT \cite{wei,liu}. This reentrant behavior does not occur at low temperature and that was found in a classical approach (MFA) 
that does not give the correct description of the ground-state phase diagram, with the presence of reentrant behavior near $h=h_{c}$ and the quantum critical point $\delta_
{c}/z=1.00$ is independent of the coordination number $z$ \cite{ovchinnikov,neto04}.

\section{Conclusions}

We have used the effective-field theory with correlation in single site cluster to obtain the state equations of the TIM antiferromagnetic. We study the phase diagrams in $h-T$ 
and $\delta-T$ planes for the square ($z=4$) lattice. The transverse field $\delta$ has important effects on the phase diagrams because it destroys the long-range order of the 
system. The results show that for a given $\delta$, the critical longitudinal magnetic field $h$ decreases with increasing temperature. In the phase diagrams there are no reentrant 
phenomena. Our results (EFT) are consistent with second-order transitions from the antiferromagnetic phase to paramagnetic phase at null transverse field. The qualitative results 
for the phase diagrams can be obtained by using more reliable methods such as quantum Monte Carlo simulation and renormalization group approaches. Furthermore, the investigations 
of this three-dimensional model are expected to show many characteristic phenomena. They will be discussed in the work.

\textbf{ACKNOWLEDGEMENT}

We thank Dr. Mircea D. Galiceanu and Dr. Octavio R. Salmon of the Universidade Federal do Amazonas for valuable discussions. This work was partially supported 
by CNPq (Edital Universal) and FAPEAM (Programa Primeiros Projetos - PPP) (Brazilian Research Agencies).


\vspace{20.0cm}


\vspace{0.01cm}


\begin{thebibliography}{9}

\bibitem{deGennes} P. G. de Gennes, \emph{Solid State Commun.} \textbf{1}, (1963) 132.

\bibitem{stinchcombe} R. B. Stinchcombe, \emph{J. Phys. C} \textbf{6}, (1973) 2459.

\bibitem{samara} G. Samara, \emph{Ferroeletrics} \textbf{5}, (1973) 25.

\bibitem{blinc} R. Blinc and B. Zeks, \emph{Adv. Phys.} \textbf{21}, (1972) 693.

\bibitem{1elliott} R. J. Elliott and A. P. Young, \emph{Ferroeletrics} \textbf{7}, (1974) 23.

\bibitem{fisher} D. S. Fisher, \emph{Phys. Rev. Lett.} \textbf{69}, (1992) 534.

\bibitem{saber} A. Saber, et al., \emph{J. Phys. Condens. Matter} \textbf{11}, (1999) 2087.

\bibitem{sarmento1} E. F. Sarmento, I. P. Fittipaldi, and T. Kaneyoshi, \emph{J. Magn. Magn. Mater.} \textbf{104-107}, (1992) 233.

\bibitem{bouziane} T. Bouziane and M. Saber, \emph{J. Magn. Magn. Mater.} \textbf{321}, (2009) 17.

\bibitem{saxena1} V. K. Saxena, \emph{Phys. Lett. A} \textbf{90}, (1982) 71.

\bibitem{saxena2} V. K. Saxena, \emph{Phys. Rev. B} \textbf{27}, (1983) 6884.

\bibitem{canko} O. Canko, E. Albayrak, and M. Keskin, \emph{J. Magn. Magn. Mater.} \textbf{294}, (2005) 63.

\bibitem{creswick} R. J. Creswick, et al., \emph{Phys. Rev. B} \textbf{38}, (1988) 4712.

\bibitem{pfeuty} P. Pfeuty, \emph{Ann. Phys.} (N. Y.) \textbf{57}, (1970) 79.

\bibitem{chakrabarti} B. K. Chakrabarti, A. Dutta, and P. Sen, \emph{Quantum Ising Models and Phase Transitions in Transverse Ising Sistems}, 
Lectures Notes in Physics, vol. M41, Springer-Verlag, Berlim, 1996.

\bibitem{sachdev} S. Sachdev, \emph{Quantum Phase Transitions}, Cambridge Univ. Press, Cambridge, 1999.

\bibitem{vettier} C. Vettier, H. L. Alberts, and D. Bloch, \emph{Phys. Rev. Lett.} \textbf{31}, (1973) 1414.

\bibitem{sughi1} S. Salem-Sugui, W. A. Ortiz, and A. D. Alvarenga, \emph{Phys. Rev. B} \textbf{40}, (1989) 2589.

\bibitem{sughi2} S. Salem-Sugui and W. A. Ortiz, \emph{Phys. Rev. B} \textbf{43}, (1991) 5784.

\bibitem{suzuki} M. Suzuki, \emph{Prog. Theor. Phys.} \textbf{56}, (1976) 1454.

\bibitem{sarmento2} E. F. Sarmento and T. Kaneyoshi, \emph{Phys. Rev. B} \textbf{48}, (1993) 3232.

\bibitem{aharony} A. Aharony, \emph{Phys. Rev. B} \textbf{41}, (1978) 3318.

\bibitem{1ma} Y. Q. Ma and Z. Y. Li, \emph{Phys. Rev. B} \textbf{41}, (1990) 11392.

\bibitem{2ma} Y. Q. Ma, et al., \emph{Phys. Rev. B} \textbf{44}, (1991) 2373.

\bibitem{kaneyoshi} T. Kaneyoshi, et al. \emph{Phys. Rev. B} \textbf{48}, (1993-I) 250.

\bibitem{weng} X. M. Weng and Z. Y. Li, \emph{Phys. Rev. B} \textbf{53}, (1996) 12142.

\bibitem{cassol} T. F. Cassol, et al., \emph{Phys. Lett. A} \textbf{160}, (1991) 518.

\bibitem{dosSantos} R. R. dos Santos, \emph{J. Phys. C} \textbf{15}, (1982) 3141.

\bibitem{2elliott} R. J. Elliott and I. D. Saville \emph{J. Phys. C} \textbf{7}, (1974) 3145.

\bibitem{hk} R. Honmura and T. Kaneyoshi \emph{J. Phys. C} \textbf{12}, (1979) 3979.

\bibitem{barreto} F. C. S\'a Barreto, I. P. Fittipaldi, and B. Zeks, \emph{Ferroelectrics} \textbf{39}, (1981) 1103.

\bibitem{sen} P. Sen, \emph{Phys. Rev. E} \textbf{63}, (2001) 16112.

\bibitem{ovchinnikov} A. A. Ovchinnikov, et al., \emph{Phys. Rev. B} \textbf{68}, (2003) 214406.

\bibitem{neto04} Minos A. Neto and J. Ricardo de Sousa, \emph{Phys. Lett. A} \textbf{330}, (2004) 322.

\bibitem{araujo} I. J. Ara\'ujo, J. Cabral Neto, and J. Ricardo de Sousa, \emph{Physica A} \textbf{260}, (1998) 150.

\bibitem{mc} R. M. Stratt, \emph{Phys. Rev. B} \textbf{33}, (1986) 1921.

\bibitem{dm} M. S. L. du Croo de Jongh and J. M. J. Leeuwen, \emph{Phys. Rev. B} \textbf{57}, (1998) 8494.

\bibitem{efrg} Q. Jiang and X. F. Jiang, \emph{Phys. Lett. A} \textbf{224}, (1997) 196.

\bibitem{mc2} O. Canko and E. Albayrak, \emph{Phys. Lett. A} \textbf{340}, (2005) 18.

\bibitem{wei} G. Wei, et al., \emph{Phys. Rev. B} \textbf{76}, (2007) 054402.
 
\bibitem{liu} J. Liu, et al., \emph{Phys. Stat. Sol. B} \textbf{244}, (2007) 3352.

\end{thebibliography}
\end{document}